%% file: main.tex
\documentclass{article}
\usepackage[utf8]{inputenc}
\usepackage{graphicx}
\usepackage{tabularx}
\def\be{\begin{equation}}
\def\ee{\end{equation}}
\def\bea{\begin{eqnarray}}
\def\eea{\end{eqnarray}}

\usepackage{caption}
\usepackage{lineno}
\usepackage{float}

\title{The simulations chain of the MURAVES experiment}

\author{M. Moussawi$^{1}$ \and S. Basnet$^{1}$ \and L. Bonechi$^{5}$ \and L. Cimmino$^{3,4}$ \and R. D’Alessandro$^{5,6}$ \and M. D’Errico$^{3,4}$ \and A. Giammanco$^{1}$ \and R. Karnam$^{1}$ \and G. Macedonio$^{7}$ \and C. Rendon$^{2}$ \and A. Samalan$^{2}$ \and G. Saracino$^{3,4}$ \and M. Tytgat $^{2}$ }

\date{$^1$Centre for Cosmology, Particle Physics and Phenomenology (CP3), Université catholique de Louvain, Louvain-la-Neuve, Belgium\\
$^2$Department of Physics and Astronomy, Ghent University, Ghent, Belgium \\
$^3$University of Naples Federico II, Napoli, Italy\\
$^4$INFN sez. di Napoli, Naples, Italy\\
$^5$INFN sez. di Firenze, Florence, Italy\\
$^6$University of Florence, Florence, Italy\\
$^7$INGV, Osservatorio Vesuviano, Naples, Italy\\
     [3ex] \today}

\begin{document}

\maketitle

{\raggedleft CP3-22-08 \\
}

\begin{abstract}
\input{abstract}
\\
\\
{\it Proceedings of the International Workshop on Cosmic-Ray Muography (Muography2021), 24-26 Nov. 2022 at Ghent, Belgium. Submitted to Journal for Advanced Instrumentation in Science.}
\end{abstract}

\section{Introduction}
\input{intro}

\input{bulk_text}

\section*{Acknowledgements}
\input{acknowledgements}

\bibliographystyle{unsrt}
\bibliography{muraves}

\end{document}

%% file: abstract.tex
The MUon RAdiography of VESuvius (MURAVES) project aims at the study of the summital cone of Mt. Vesuvius, an active and hazardous volcano near Naples, Italy. 
A detailed Monte Carlo simulation framework is necessary in order to investigate the effects of the experimental constraints and to perform comparisons with the actual observations. Our Monte Carlo setup combines a variety of Monte Carlo programs that address different aspects of cosmic muon simulation, from muon generation in the Earth's upper atmosphere to the response of the detector, including the interactions with the material of the volcano. 
We will elaborate on the rationale for our technical choices, including trade-off between speed and accuracy, and on the lessons learned, which are of general interest for similar use cases in muon radiography.

%% file: intro.tex
\label{sec:intro}

Muon radiography is an imaging method that uses natural cosmic ray muons to characterize the density distribution of large scale natural or human-made objects, complementary to conventional imaging techniques~\cite{Bonechi:2019ckl}. Imaging is performed from the measurements of the absorption profiles of muons as they pass through matter~\cite{Lechmann:2021brn}. The MUon RAdiography of VESuvius (MURAVES) project~\cite{MURAVES} aims at the study of the inner structure of the upper part of the Mt. Vesuvius, an active volcano near Naples, Italy. The main expected outcome of MURAVES will be a muography of Mt. Vesuvius, i.e. to obtain a 2D image of the density structure by measuring the attenuation of the muons flux trough the volcano cone. 
By combining sufficiently precise data from muography with those from gravimetric and seismic measurement campaigns, it may become possible to determine the stratigraphy of the lava plug at the bottom of the Vesuvius crater, in order to infer potential eruption pathways.

The detection setup of MURAVES consists of three identical and independent tracking hodoscopes, constituted of scintillator bars coupled to SiPMs, which are already installed on Mt. Vesuvius and fully operational. Each hodoscope includes four tracking stations, each with a surface of roughly 1 m$^2$, and a thick lead wall between the 3rd and the 4th station which is used for muon energy discrimination, as explained later in this document.

Comparing data with Monte Carlo (MC) simulations is very important for the imaging of a target. A popular approach in muography is to compare various density hypotheses in MC with the observed transmission map, seeking the simulated hypothesis that provides the best fit to real data. In other types of studies, one is interested in anomalies with respect to the expected map from MC. In all those cases, various effects can bias the expectations, thus it is important to quantify their impact by testing various modeling assumptions in MC.

\begin{figure}[!ht]
    \centering 
    \includegraphics[width=\textwidth]{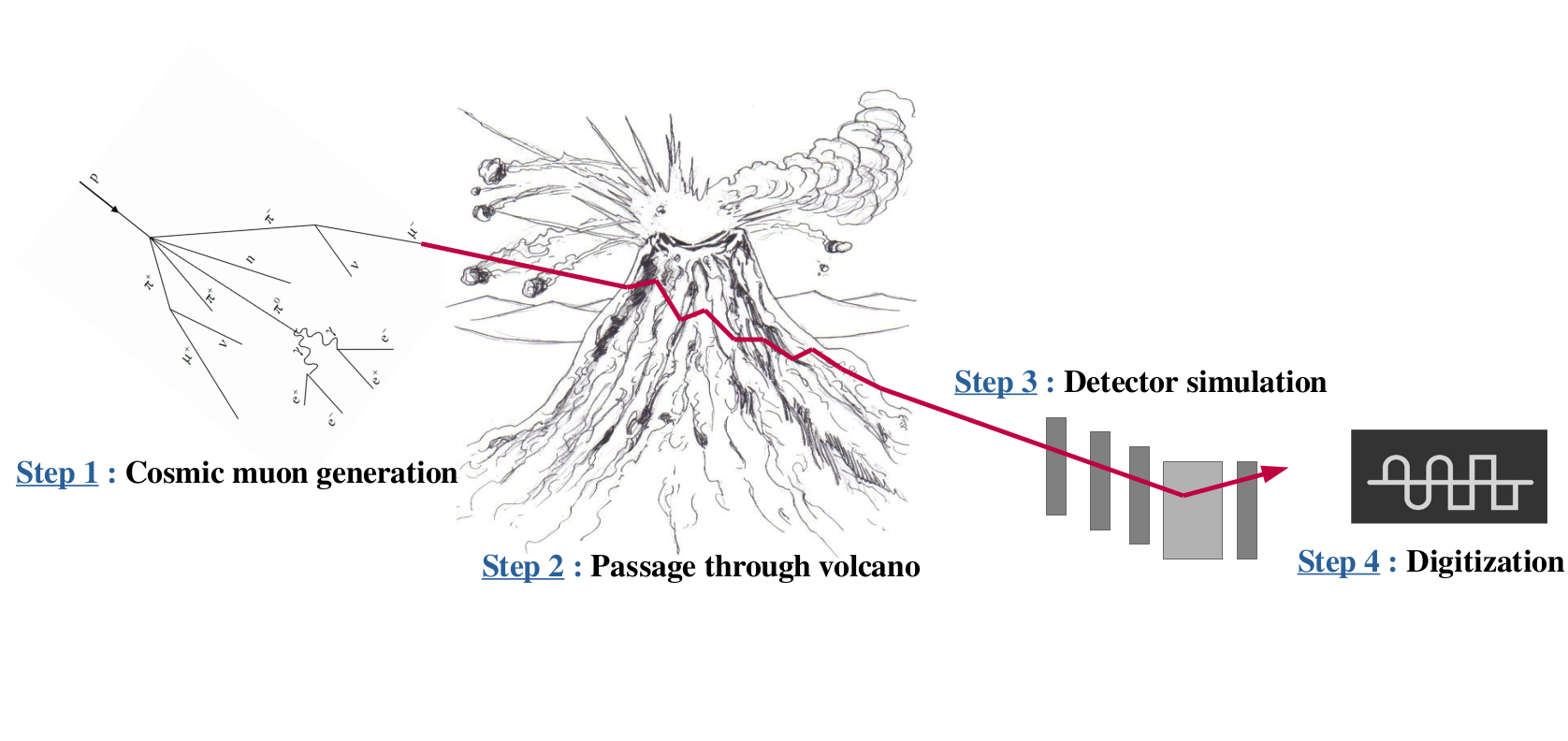}
    \caption{\label{fig:chain}A schematic representation of the MURAVES simulation chain.}
\end{figure}

In this paper, we report on the development of a MC simulation chain based on the schematic steps illustrated in Figure ~\ref{fig:chain}, and we present a series of simulation studies that are being conducted to investigate the effects of the experimental constraints and to perform comparisons with the actual observations. 
Our simulation chain starts from the generation of cosmic muons with realistic angular and momentum distributions. Several tools for this task are available from {\it ab initio} simulations that simulate the development of the entire cosmic shower in the atmosphere (e.g. CORSIKA~\cite{CORSIKA}), to parametric simulations (such as CRY~\cite{CRY} and EcoMug~\cite{EcoMug}) i.e. Monte Carlo programs that sample muon properties from pre-defined distributions. 
For the simulation of muon traversal through the volcano, we are investigating PUMAS~\cite{PUMAS}, a "backward MC" designed to optimize computational efficiency by only simulating the muons whose trajectories actually reach the detectors.   
A detailed simulation of the detector response and of its geometry becomes more and more important as the accumulation of statistics makes the muography of the volcano more precise; for that, our work-horse is 
GEANT4~\cite{GEANT4}, a multipurpose tool very popular in particle, astroparticle, nuclear and medical physics. 
To mimic the real data, GEANT4 raw hits are converted to clusters through a simulated digitization. After application of the tracking algorithm, we compare a few high-level quantities with data.

%% file: bulk_text.tex
\section{Cosmic-ray muon generation}
\label{sec:Cosmicgenerator}

The first simulation step is the generation of muons, as well as of background particles such as electrons, positrons, hadrons, according to realistic models.

We examined and compared three Monte Carlo tools: CRY~\cite{CRY}, CORSIKA~\cite{CORSIKA} and EcoMug~\cite{EcoMug}. CRY and EcoMug are parametric simulations, while CORSIKA simulates the evolution of extensive air shower in the atmosphere step-by-step. 
EcoMug allows the generation of cosmic muons according to user-defined parametrizations of their differential flux, and a default one is provided based on data from the ADAMO experiment~\cite{adamo}. 
On the other hand, CRY's parameterization, based on MCNPX simulation~\cite{MCNPX} takes into account geomagnetic effects on the cosmic flux depending on time (solar cycle), latitude and altitude (for the latter, CRY provides three options: 0, 2100, 11300~m above sea level). 
Instead, in CORSIKA the user can choose across various models of hadronic interaction at low and high collision energies. 

Both absorption and scattering of the muons depend strongly on the particle momentum. 
Of particular importance in this context is the very high-energy end of the spectrum, as only the most energetic muons are able to pass through the deepest and most interesting regions of a volcano; e.g., it takes a bit less than 10~TeV for a muon to pass through 2.5~km of rock. 
In Ref.~\cite{Muraves_simulation}, we compared the muon flux as a function of kinetic energy for CRY and five different combination of low and high energy hadronic interaction models available in CORSIKA in the range between 1~MeV and 10~TeV, finding them all in agreement concerning the peak position and the general shape but with a certain spread in predictions at high energy. 
These differences need to be taken into account as systematic uncertainties in the MURAVES measurements. To that end, the possibility in CORSIKA to vary the hadronic model can be exploited for assessing a physics-motivated modeling uncertainty.

\begin{figure}[!ht]
 \centering 
 \includegraphics[width=1\textwidth]{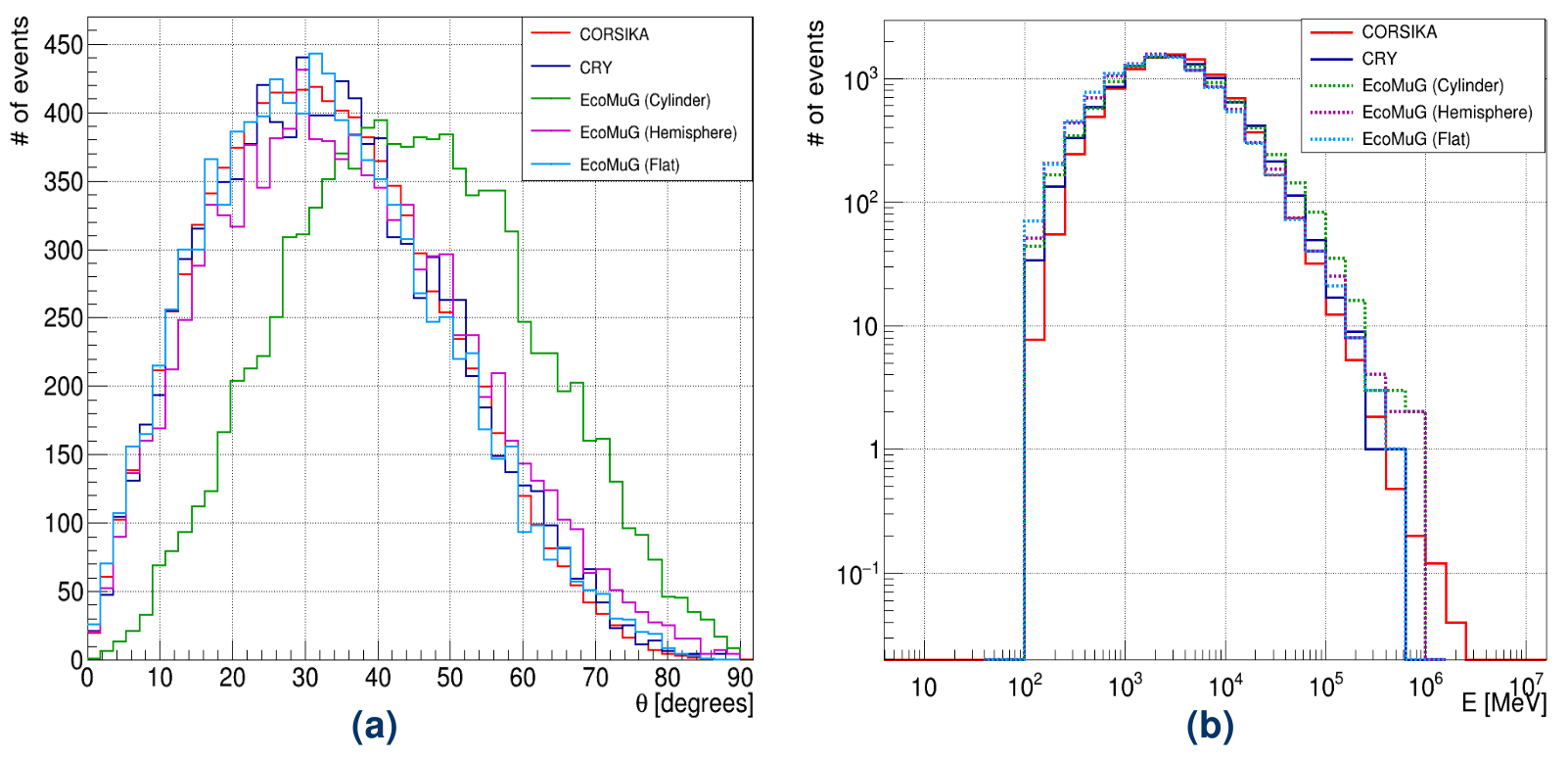}
 \caption{\label{fig:Generator}Distributions of zenith angle $\theta$ (a) and kinetic energy $E$ (b) generated using CRY, CORSIKA and three EcoMug modes (corresponding to flat, cylindrical and hemispherical generation planes).}
 \end{figure}

In a particle generator, the user has to define the surface from where the particles start their trajectory, which is called the ``generation surface''. 
While most cosmic-particle generators, including CRY and CORSIKA, are limited to flat generation surfaces, EcoMug allows to shoot muons also from cylindrical and hemispherical surfaces, ensuring the correct angular and momentum distributions. 
Figure \ref{fig:Generator} shows the distribution of zenith angle $\theta$ (a) and kinetic energy $E$ (b) of muons simulated using CRY, CORSIKA (with the DPMJET~\cite{DPMJET} and GHEISHA~\cite{GHEISHA} hadronic interaction models used respectively for high and low energy collisions in the atmosphere~\cite{Heck:2004rq}) and the three EcoMug modes.
It can seen that the three generators yield consistent distributions for $E$; in $\theta$ all generators are consistent when used with a flat generation surface, while the cylindrical configuration of EcoMug shows a substantial angular bias. Intuitively, this is explained by recalling that $\theta \sim 0^\circ$ (very vertical muons) can never happen in this configuration as it lacks the top of the cylinder.

An important practical consideration is the speed of execution, for which there are huge differences: we estimated that to generate 10$^5$ muons on a standard CPU, EcoMug employs O(sec) (depending on the mode: flat is faster, hemisphere is slower, as thoroughly studied in Ref.~\cite{EcoMug}), CRY O(min), while CORSIKA takes O(hours).

Table~\ref{tab:generators} compares the relative merits of the three generators examined, from the point of view of what is important for a muon radiography experiment similar to ours. 
For the purpose of MURAVES data analysis, we chose to use CRY as main generator, while also maintaining CORSIKA and EcoMug (with its flat-surface option) for the purpose of estimating systematic uncertainties related to physics modeling.  
Both CRY and EcoMug can easily be interfaced to GEANT4~\cite{GEANT4}, meaning that by compiling the appropriate libraries it is possible to run muon generation and detector response simulation in the same job. 
Running GEANT4 on CORSIKA events requires, instead, to save the generator output in HepMC format~\cite{HepMC}, which complicates the storage logistics. 
Additional considerations, although less crucial, include the ability to simulate non-muon backgrounds from cosmic showers (e.g., electrons, positrons, hadrons) and multiple muons from the same shower (hence detected simultaneously in the same hodoscope, leading to rejection by the current tracking algorithm). Both features are naturally included in CORSIKA, as it develops the full showers. CRY contains parametrizations of all particles including backgrounds, based on MCNPX~\cite{MCNPX} and can handle multiple particles per event. EcoMug is not designed for multiple muons or for backgrounds, and examples are provided only for muons; however, we implemented and tested a technical solution for simulating any kind of particle, after private communication with the authors, that might be of interest for future studies.

\begin{table}[h]
\begin{tabular}{l|l|l|l}
\textbf{Generator} & CORSIKA & CRY & EcoMug \\ \hline
\textbf{Speed (10$^5$ muons)} & O(hours) & O(min) & O(sec) \\
& & & \\\hline
 & Ab initio, & Immutable, & User-defined,  \\
\textbf{Modeling} & several models & from MC~\cite{MCNPX} & default from \cite{adamo} \\
 & available & & \\
& & & \\ \hline
\textbf{Accuracy} & State of the art & Approximate & Approximate \\
& & & \\ \hline
\textbf{Use with GEANT4} & Complex & Easy & Easy \\
& & & \\ \hline
 & Hadronic & Time & Alternative \\
\textbf{Systematics} & interaction & dependence & parameterizations\\
 & models & (solar cycle) & \\

\end{tabular}
\caption{Qualitative comparison of the features of CORSIKA~\cite{CORSIKA}, CRY~\cite{CRY} and EcoMug~\cite{EcoMug} that are of particular interest for muon radiography of a volcano.}
\label{tab:generators}
\end{table}

\section{Passage of muons through Vesuvius}
\label{sec:volcano}

The probability of a muon surviving the crossing of a certain amount of material is determined by the opacity of the latter (defined as the integrated matter density $\rho$ over the path length, i.e. $\omega \equiv \int_{entry}^{exit} \rho(x) dx$) and the initial energy of the muon. The measurement of the transmitted flux of atmospheric muons through a target in different directions from a given point of view (the position of the muon detector) provides the two-dimensional projective measurement of the target, and the average density ($\bar \rho$) of the material of this target can be calculated by knowing the opacity using the equation $\omega = L\bar \rho$, where $L$ is the total muon path length within the material. The path lengths corresponding to different zenith and azimuth angles of arrival of the muons, across the Vesuvius crater, are evaluated using a Digital Terrain Model (DTM) with 1 m resolution, derived using the data from ~\cite{vilardo2013morphometry}. 

The computation of the expected muon flux through Mt. Vesuvius 
is conducted using PUMAS~\cite{PUMAS}, a muon transport library based on a Backward Monte Carlo (BMC) technique. 
The concept of a BMC can be concisely described as ``a MC run backwards'', i.e. from the detector to the sky. 
A BMC deals with the exact same problem as a forward MC, but it allows to generate only the final states that are actually observable, thus reducing the computation time considerably. 
In our use case, it is important to simulate the scattering of the muons when they pass through the mountain. In locations which are close to the mountain, we can expect considerable amount of low energy muons suffering multiple scattering near the surface and being deflected into the acceptance of the detector at a fake direction. 

The choice of the optimal observation point where to install the MURAVES detector was based on PUMAS (in collaboration with its authors), integrated with TURTLE~\cite{TURTLE} to extract the path lengths from the DTM. Various potential installation points, preselected based on logistical considerations, were studied with PUMAS and the muon flux was simulated given the thickness and average density of the material. Among them, one was chosen based on its clear view of Mt. Vesuvius and on the signal purity expected over most directions. In this context, purity is defined as the ratio of the muon flux without scattering over the muon flux with scattering, and it is used as a figure of merit; low purity means a large contamination from low-energy muons whose entry and exit directions are different due to multiple scattering, hence blurring the muography resolution.

 \begin{figure}[!ht]
 \centering 
 \includegraphics[width=1\textwidth]{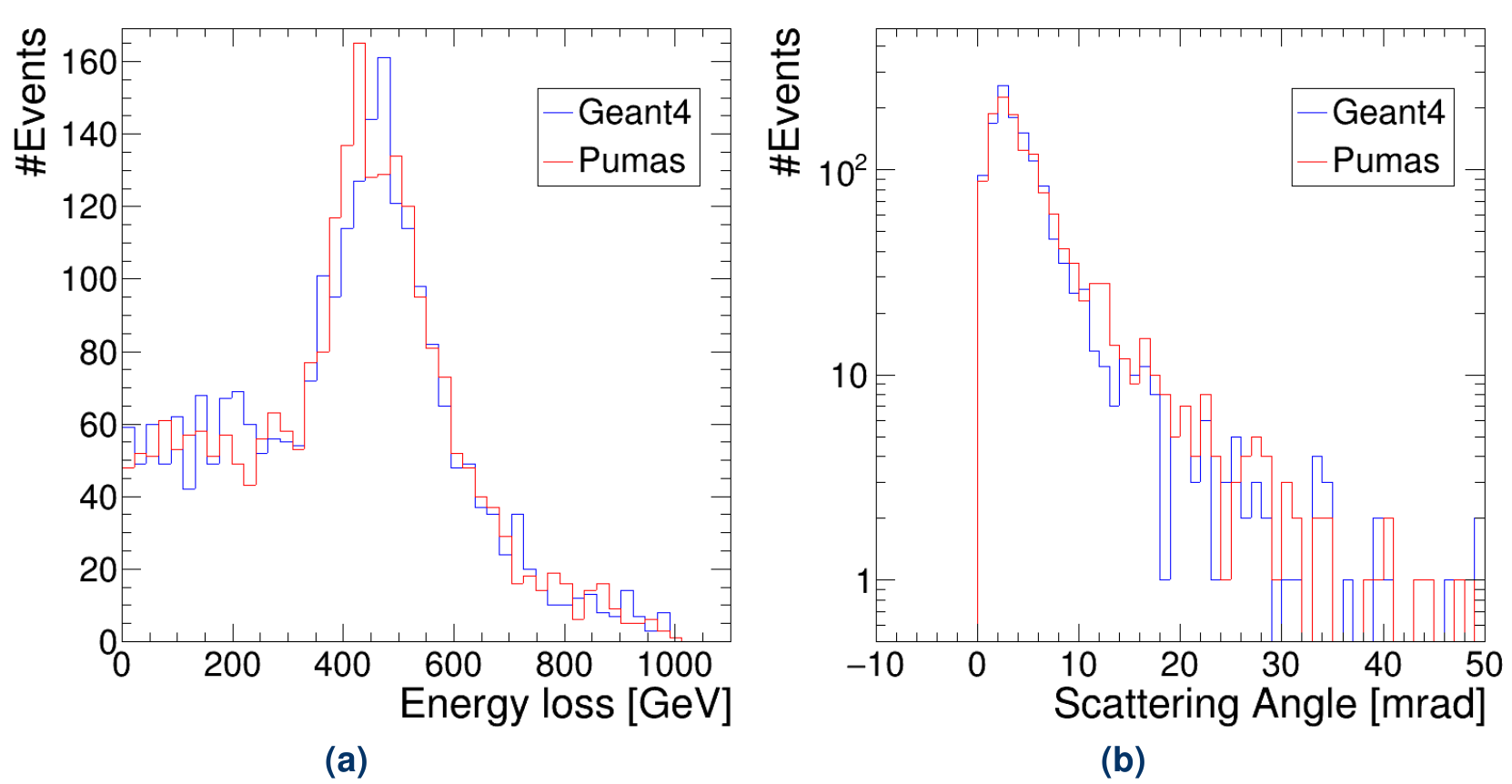}
 \caption{\label{fig:geant-vs-pumas}Distributions of energy loss (a) and scattering angle (b) of 0-1000 GeV muons in 500 m of standard rock generated using GEANT4 and PUMAS.} 
\end{figure}

The accurate simulation of muon transport through the volcano is also important in order to translate the measured transmission map into an opacity map. 
The larger computation efficiency of PUMAS with respect to forward MC programs makes it a very appealing option, but the relative novelty of the BMC approach imposes the need to validate it against a ``golden standard''.
For that reason, we studied the energy loss and scattering of muons in the rock using PUMAS and GEANT4, and the results of the two simulations were compared. Muons with a flat energy distribution ranging from 0 GeV to 1000 GeV were simulated and the muon path length was set to 500 m in standard rock. Figure \ref{fig:geant-vs-pumas} shows the distribution of energy loss (a) and scattering angle (b) evaluated using PUMAS (red line) and GEANT4 (blue line). We conclude that the two programs give reasonably close results.

\section{Detector simulation}
\label{sec:detctsim}

GEANT4~\cite{GEANT4} is the most popular MC particle transport simulation in particle and nuclear physics, as well as medical and space applications. 
For the MURAVES simulation studies, we built a detailed model of a MURAVES hodoscope in GEANT4, interfaced with CRY as source for all signal and background particles. 
For specific studies, single muons have been simulated at specific energies and incident positions using GEANT4 "particle gun" generator or General Particle Source class. We also produced samples where GEANT4 was interfaced with EcoMug, and where the input was taken from the HepMC files previously produced with CORSIKA.

\begin{figure}[!ht]
\centering 
\includegraphics[width=0.6\textwidth]{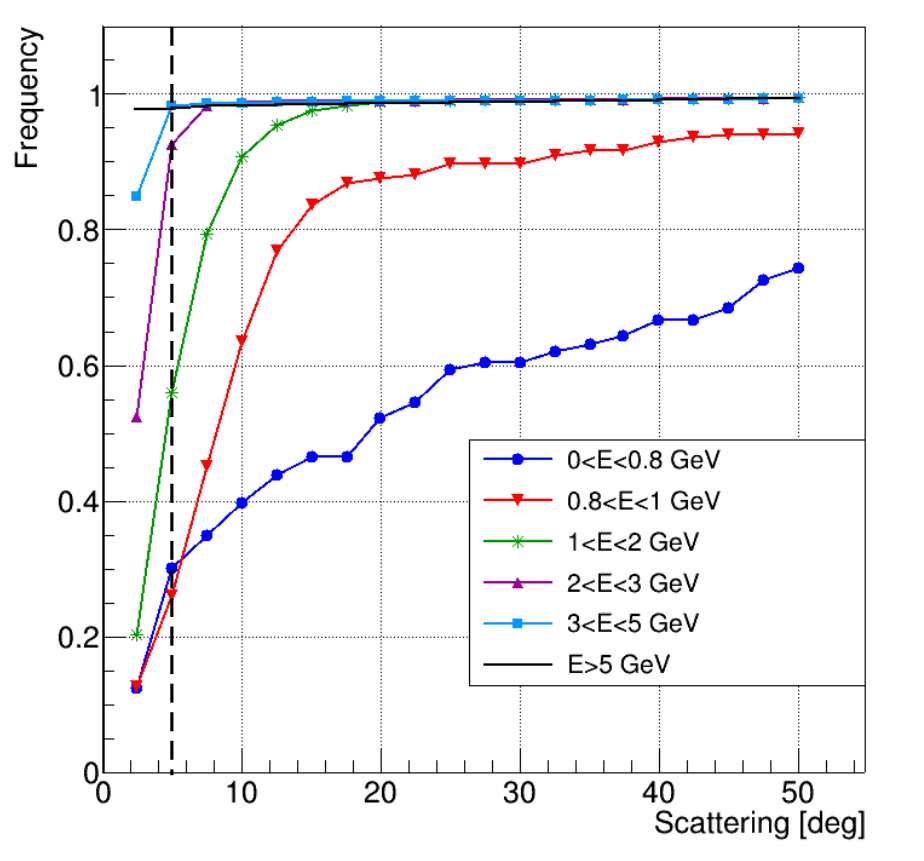}
\caption{\label{fig:eff-scattering}Efficiency of various possible upper cuts on the scattering angle for different ranges of incoming muons energy.} 
\end{figure}

The MURAVES setup consists of three identical and independent tracking hodoscopes, constituted of scintillator bars coupled to Silicon Photo-Multipliers (SiPMs). Two of them are positioned in ``forward-looking mode'' to take Mt. Vesuvius data, while one is installed in reverse orientation, to acquire a free-sky control sample. The MC simulation is applied to the three hodoscopes, considered as strictly identical.

Each hodoscope includes four tracking stations. A thick lead wall of thickness 60 cm between the 3rd and the 4th stations acts as 'minimum muon energy' selector with a turn-on point of 0.9~GeV~\cite{Muraves_simulation}, and also reduces very effectively the background from electrons, positrons or hadrons. Each station has two planes (X and Y) with 64 adjacent scintillator bars (divided into two modules) housed inside an aluminium container and of 1~m$^2$ sensitive area placed orthogonal to each other to extract the bi-dimensional information of the muon hits. The scintillator bars are of triangular cross-section with base width of 33 mm and height 17 mm, and alternate upward and downward pointing tips. 

We used GEANT4 to guide the choice of a possible analysis selection step based on the scattering induced by the lead wall between the 3rd and the 4th stations. 
The purpose is to further reduce the fraction of low-energy muons, after the passive filtering provided by the absorption in the lead wall, while keeping as many as possible of the high-energy ones, which carry the cleanest information about the depths of rock traversed. 
To perform this study, we generated 40 million muons with CRY, in free-sky conditions, and we selected the events containing a signal in all hodoscope stations. 
GEANT4 gives access to the scattering points within the lead wall; however, those are unobservable in real data, therefore we approximate the actual scattering angle with the angle between the vector connecting the 1st- and 3rd-station hits, and the vector connecting the 3rd- and the 4th-station hits. This approximation mimics the practical requirement that can be applied on tracks in real data. 
Finally, we counted the frequency at which the events survive an upper cut on this angle. 
Figure~\ref{fig:eff-scattering} shows the outcome of this study, for muons belonging to various incoming energy categories. The horizontal axis indicates the upper cut on the scattering angle (in steps of $2.5^\circ$), while the vertical axis shows the frequency of survival of a muon, defined as the fraction of muons passing this cut, among those that give hits in all stations (hence, that were not absorbed in the lead wall). 
It can be seen that even a very restrictive cut of $\le 5^\circ$ preserves the vast majority of muons above 2~GeV (better than 90\% between 2 and 3~GeV, and almost 100\% above 3~GeV), while removing more than 40\% of the muons between 1 and 2~GeV and about 70\% of the sub-GeV muons that pass the lead wall, whose angular distribution is very broad. 
 
\section{Digitization, clustering and tracking}
\label{sec:digit}

The output of GEANT4 contains "hits", corresponding to each interaction of a particle with the sensitive detector material (scintillator bars) for each simulation step. Hits are characterized by 3D position, time, energy deposition, and identity of the particle involved. 
In GEANT4, when a muon passes through a detector plane, it typically produces multiple hits in the scintillator bars, while we get only a single signal per bar in the real case. 

Our next simulation step is the "digitization" of these GEANT4 raw hits, i.e. the quantisation of both their positions and energy deposits. 
Energy deposits are first summed per scintillator bar, and then converted to number of photoelectrons (nPE) using a conversion factor of 10~nPE every 1.4~MeV, based on the comparison of the peak deposited energies observed in real and simulated data, respectively. 
At this point we have a single value of the energy deposit per bar, expressed in proper units, and we can apply the clustering algorithm. A cluster is a collection of adjacent bars whose signals pass a certain threshold; clusters may also be constituted of a single bar, in which case however the signal must pass a stricter threshold. 
When a cluster contains more than one bar, its position is calculated from their barycenter, i.e. the weighted average of the positions of the centers of the bars involved, where the weights are given by the corresponding signals, measured in number of photoelectrons.
The same clustering algorithm and thresholds are applied in MC as in real data.

\begin{figure}[H]
\centering
\begin{minipage}{13pc}
\includegraphics[width=13pc]{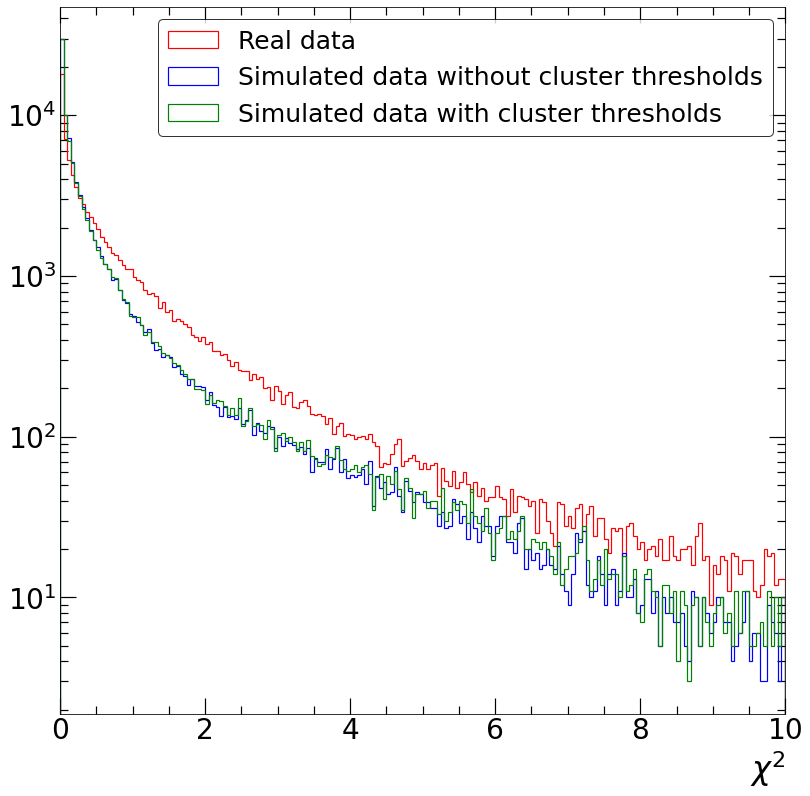}
\caption{Comparison of the $\chi^2$ of the tracks in real data and simulated data (the latter with and without clustering thresholds).}
\label{chi_sqr}
\end{minipage}\hspace{1pc}
\begin{minipage}{13pc}
\includegraphics[width=13pc]{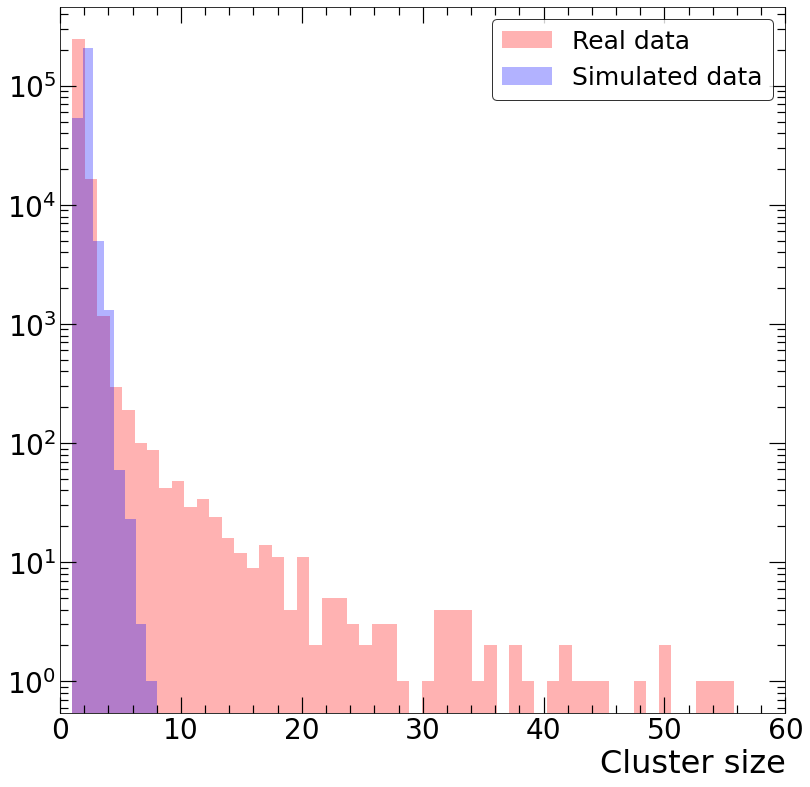}
\caption{Comparison of the sizes of the clusters involved in the selected tracks ($\chi^2<$5) for real and simulated data.}
\label{clsz}
\end{minipage} 
\end{figure}

Finally, the simulated clusters are fed into the tracking algorithm, which again is rigorously applied as in real data. 
The tracking is performed for X planes and Y planes independently. To validate the full simulation chain, 
the $\chi^2$ values of the simulated tracks are compared with those of the real data from a free-sky calibration run.
For this, 10$^5$ muon events were simulated, all crossing at least the first three planes of the detector, and the full reconstruction chain was applied. 
The result is shown in Figure \ref{chi_sqr}, which also includes an alternative MC dataset where no thresholds are applied in the clustering algorithm. 
Although the trend is qualitatively similar between data and MC, the latter is significantly too optimistic, i.e. it has a lower tail than real data. This difference is found to be unrelated to the applied cluster thresholds, as demonstrated by the similarity of the MC distributions with and without thresholds. 

A possible insight in the source of this discrepancy is offered by Figure \ref{clsz}, which shows the comparison of the size of the clusters involved in the tracks selected with the standard cut $\chi^2 \le 5$. As per the results, it is clear that MC is currently unable to simulate the long tail that we observe in the real data and more investigations are needed in order to find out which unmodeled effects are the culprits; in fact, our simulation chain does not consider (yet) several nuisances, such as dark noise and various front-end electronics effects. 
The discrepancy in $\chi^2$ distribution could be solved either by emulating such effects or by further cleaning the real data by more optimal selections at cluster level.

\section{Conclusions}
 We presented a complete Monte Carlo simulation chain for the MURAVES experiment. 
For the generation of cosmic muons, a comparison study between three different
particle generators (CRY, CORSIKA and EcoMug) was conducted, and it was
decided to use CRY as main generator for our analysis, with the other two kept as options for the estimation of modeling systematics.
The problem of efficient simulation of the muon passage through a large body, such as Vesuvius, is approached using PUMAS, whose accuracy we verified by comparing with GEANT4 in reference conditions. 
A detailed geometry of a MURAVES hodoscope was built using GEANT4 for the simulation of the detector response. The output of GEANT4 is digitized into a format that mimics raw data.  In order to validate the performance of the CRY-GEANT4-digitization simulation chain, we compared the $\chi^2$ distribution of real and simulated tracks, finding a difference that will deserve further investigation.

%% file: acknowledgements.tex
The authors gratefully acknowledge the precious help by Valentin Niess and Cristina Carloganu; their simulations of Mt. Vesuvius with PUMAS~\cite{PUMAS} were used at the start of the MURAVES project for the choice of the optimal location for the installation of the detectors. More recently, in private communication, Valentin Niess also kindly clarified some important physical and technical aspects in PUMAS.
We thank Germano Bonomi and Davide Pagano for useful discussions about EcoMuG~\cite{EcoMug}. 
We are also indebted to Pasquale Noli and Nicola Mori for their previous work on the simulation of MU-RAY (precursor of MURAVES)~\cite{GGS}. 
This work was partially supported by the EU Horizon 2020 Research and Innovation Programme under the Marie Sklodowska-Curie Grant Agreement No. 822185, and by the Fonds de la Recherche Scientifique - FNRS under Grants No. T.0099.19 and J.0070.21. 
